\documentclass[aps, twocolumn, letterpaper, superscriptaddress, nofootinbib]{revtex4}

\usepackage{amsmath}
\usepackage{amssymb}
\usepackage{mathrsfs}
\usepackage{xspace}
\usepackage{braket}
\usepackage{color}
\usepackage{graphicx}

\usepackage{ifpdf}
\ifpdf
\fi

\newcommand{\myRef}[1]{Ref.~\cite{#1}}
\ifx\Ref\undefined
\newcommand{\Ref}[1]{\myRef{#1}}
\else
\renewcommand{\Ref}[1]{\myRef{#1}}
\fi

\newcommand{\eq}[1]{(\ref{#1})}
\newcommand{\Eq}[1]{Eq.~(\ref{#1})}
\newcommand{\Eqs}[1]{Eqs.~(\ref{#1})}
\newcommand{\Fig}[1]{Fig.~\ref{#1}}
\newcommand{\Sec}[1]{Sec.~\ref{#1}}
\newcommand{\Refs}[1]{Refs.~\cite{#1}}

\newcommand{\eg}{{e.g.,\/}\xspace}
\newcommand{\ie}{{i.e.,\/}\xspace}

\newcommand{\pd}{\partial}
\newcommand{\dd}{\mathrm{d}}
\newcommand{\ee}{\mathrm{e}}
\newcommand{\ii}{\mathrm{i}}
\newcommand{\del}{\nabla}

\newcommand{\mc}[1]{\mathcal{#1}}

\newcommand{\msf}[1]{\mathsf{#1}}
\newcommand{\mcu}[1]{\mathscr{#1}}

\newcommand{\oper}[1]{\smash{\hat{#1}}}
\newcommand{\kpt}[1]{\kern #1 pt} 
\newcommand{\favr}[1]{\langle #1 \rangle} 
\newcommand{\qavr}[1]{\langle\!\langle #1 \rangle\!\rangle} 

\newcommand{\hH}{\mcu{H}}
\newcommand{\hV}{\mcu{V}}
\newcommand{\mH}{\mc{H}}
\newcommand{\mU}{\mc{U}} 
\newcommand{\mVN}{V} 
\newcommand{\mUN}{U} 

\begin{document}

\title{Quantum computation of nonlinear maps}

\author{I. Y. Dodin}
\affiliation{Princeton Plasma Physics Laboratory, Princeton, NJ 08543}
\affiliation{Department of Astrophysical Sciences, Princeton University, Princeton, NJ 08544}
\author{E. A. Startsev}
\affiliation{Princeton Plasma Physics Laboratory, Princeton, NJ 08543} 

\begin{abstract}
Quantum algorithms for computing classical nonlinear maps are widely known for toy problems but might not suit potential applications to realistic physics simulations. Here, we propose how to compute a general differentiable invertible nonlinear map on a quantum computer using only linear unitary operations. The price of this universality is that the original map is represented adequately only on a finite number of iterations. More iterations produce spurious echos, which are unavoidable in any finite unitary emulation of generic non-conservative dynamics. Our work is intended as the first survey of these issues and possible ways to overcome them in the future. We propose how to monitor spurious echos via auxiliary measurements, and we illustrate our results with numerical simulations.
\end{abstract}

\maketitle



\section{Introduction} There is a growing interest in applications of quantum computing to modeling classical systems \cite{ref:engel19, ref:parker20, ref:costa19, ref:scherer17, ref:montanaro16, tex:myqc} and nonlinear classical systems in particular \cite{tex:myqc, ref:joseph20, tex:liu20, tex:engel20, tex:lloyd20, tex:shi20, ref:lubasch20, tex:leyton08}. Closely related to this problem is quantum computation of nonlinear classical maps $x \mapsto X(x)$ that are generally \textit{not} near-identity (unlike, say, maps involved in numerical solvers of differential equations). Such maps emerge in stroboscopic description of physical systems \cite{book:lichtenberg}, and they also enter classic optimization algorithms, such as Newton’s algorithm and the gradient-descent algorithm. It is of interest if computation of these maps can be speeded up using quantum computers (QCs). Before looking into applications, though, a more fundamental question must be addressed first: what does it take to calculate a nonlinear function on a QC in general?

Quantum algorithms for computing nonlinear maps have been widely discussed in literature \cite{ref:georgeot08, ref:frahm04, ref:levi04, ref:levi03, ref:georgeot01, ref:georgeot01b, ref:benenti01, ref:schack98} but only for toy maps that are: (i)~area-preserving (AP) and (ii)~can be explicitly quantized, \ie represented as linear maps for a state function $\psi \mapsto \ee^{-\ii\oper{\mH}}\psi$ with known Hermitian $\oper{\mH}$. Such maps may be of interest for basic research, but they are not representative of those anticipated in potential applications. For example, a gradient-descent algorithm for finding a local minimum of a function $f$ involves iterative applications of the map
\begin{gather}
X(x) = x - \eta \del f(x),
\end{gather}
where $\eta$ is a constant. This map is non-AP and cannot be easily quantized in the general case. Quantum implementation of such maps requires approaches very different from those commonly considered in literature.

Here, we propose how to implement a general differentiable invertible map $X$ on a QC. Our algorithm allows calculating individual nonlinear ``trajectories'' $x(t)$ generated by $t$ iterative applications of $X$. Our scheme represents the original map adequately on a finite number of iterations, and it is universal; for example, it can be applied to minimization algorithms such as the steepest-descent algorithms. Below, we outline the general theory and illustrate its application (on a classical computer) using a simple one-dimensional map as an example. Our work is intended as a preliminary analysis of the generic, arguably unavoidable, issues that will have to be dealt with in developing nonlinear quantum algorithms for any specific applications in the future.

Our paper is organized as follows. In \Sec{sec:prob}, we formulate the general problem. In \Sec{sec:unit}, we explain how a generic nonlinear map can be represented as a linear unitary map in an infinite-dimensional space. In \Sec{sec:trunc}, we discuss how to keep this map unitary (``unitarize'' the map) when that space is truncated. In \Sec{sec:basis}, we discuss how to approximate the unitarized truncated map with a sparse unitary map. In \Sec{sec:app}, we discuss potential applications, namely, calculating individual nonlinear trajectories and finding attracting fixed points of a given map. In \Sec{sec:example}, we present an example and related numerical simulations. In \Sec{sec:conc}, we summarize our main results.

\section{Problem setup} 
\label{sec:prob}

Consider a configuration space $x = \{x^1, x^2, \ldots, x^{n}\}$ and suppose some initial distribution $F$ of initial data on $x$. An invertible automorphism on this space, ${X: x \mapsto \bar{x} = X(x)}$, induces a transformation
\begin{gather}\label{eq:FF}
X: \kpt{5} F \mapsto \bar{F}(x)=\frac{F(X^{-1}(x))}{|\det\del X(X^{-1}(x))|}.
\end{gather}
Here, $X^{-1}$ is the function inverse to $X$ and $\del$ is the derivative with respect to the whole argument, so the Jacobian $\del X \equiv J(x)$ is generally a matrix with elements $J^\mu{}_\nu = \pd X^\mu/\pd x^\nu$. For AP transformations, which have $|\det J| = 1$, \Eq{eq:FF} becomes simply $\bar{F}(x) = F(X^{-1}(x))$. 

For a given $X$, it can be convenient to consider also an extended map constructed as follows. Let us introduce an auxiliary coordinate $p = \{p_1, p_2, \ldots, p_n\}$ and a $2n$-dimensional coordinate $z = \{x, p\}$. On the $z$ space, let us define a map $Z: z \mapsto \bar{z}= \{\bar{x}, \bar{p}\}$ such that
\begin{gather}\label{eq:exttr}
\bar{x} = X(x), \quad \bar{p} = [J^{-1}(x)]^{\intercal} p,
\end{gather}
where $^{\intercal}$ denotes transposition. A direct calculation shows that $\det\del Z = 1$ (and furthermore, $Z$ is by definition symplectic \cite{book:meyer17}). This means that by extending the configuration space, any non-AP map can be converted into an AP map, in which case basic theory becomes more transparent. However, introducing the extended space is not necessary for our purposes. Because of this, below we assume a general map with an arbitrary invertible Jacobian on an arbitrary $n$-dimensional space.

\section{Nonlinear map as a unitary transformation} 
\label{sec:unit}

Our scheme uses an extension of the idea that was proposed in \Refs{tex:myqc, ref:joseph20} for solving differential equations on a QC (see also \Ref{tex:engel20}). Let us introduce ``state functions'' $\bar{\Psi}\doteq \smash{\sqrt{\bar{F}}}$ and $\Psi \doteq \smash{\sqrt{F}}$ (the symbol $\doteq$ denotes definitions), which can be considered as elements of a Hilbert space $\hH$ with the inner product
\begin{gather}
\braket{f_1, f_2} = \int \dd x\,f_1^*(x)\,f_2(x).
\end{gather} 
Let us assume a countable infinite (and possibly nonorthonormal) basis $\mcu{E} = \smash{\{e_a\, |\, a = 1, 2, \ldots\}}$ on $\hH$ and decompose the state functions in $\mcu{E} $:
\begin{gather}\label{eq:decomp}
\bar{\Psi}(x) = e_a(x)\,\bar{\psi}^a,
\quad
\Psi(x) = e_a(x)\, \psi^a.
\end{gather} 
Here $\bar{\psi}^a$ and $\psi^a$ and the corresponding coefficients, and summation over repeating indices is assumed. (As usual, the distinction between the upper and lower indices can be ignored if the basis is orthonormal.) Because
\begin{gather}
\bar{\Psi}(x) = |\det \del X(X^{-1}(x)) |^{-1/2}\Psi(X^{-1}(x)),
\end{gather}
this leads to $\bar{\psi}^a = \mU[X]^a{}_b\psi^b$, where 
\begin{gather}\label{eq:U}
\mU[X]^a{}_b = \int \dd x\, \frac{e^{a*}(x)\, e_b(X^{-1}(x))}{\sqrt{|\det \del X(X^{-1}(x))|}}
\end{gather}
and $e^a$ are elements of the dual basis; \ie $\braket{e^a, e_b} = \delta^a_b$. (The notation $[X]$ denotes that $\mU^a{}_b$ is a functional evaluated on~$X$.) Using the variable transformation $x = X(\tilde{x})$, one can rewrite \Eq{eq:U} in the form
\begin{gather}\label{eq:Umn}
\mU[X]^a{}_b = \int \dd \tilde{x}\, \sqrt{|\det \del X(\tilde{x})|}\, e^{a*}(X(\tilde{x}))\,e_b(\tilde{x}),
\end{gather}
which is convenient in that it does not contain $X^{-1}$. Also notice the following. By differentiating the identity $X(X^{-1}(x)) = x$ with respect to $x$, one obtains $[\del X(X^{-1}(x))]\,\del X^{-1}(x) = 1$, which leads to
\begin{gather}\label{eq:detdet}
\det \del X(X^{-1}(x)) = 1/\det \del X^{-1}(x)
\end{gather}
for any $x$. By evaluating this at $x = X(\tilde{x})$, one can rewrite \Eq{eq:Umn} in yet another form to be used below:
\begin{gather}\label{eq:Umn2}
\mU[X]^a{}_b =\int \dd \tilde{x}\,\frac{e_b(\tilde{x})\,e^{a*}(X(\tilde{x}))}{\sqrt{|\det \del X^{-1}(X(\tilde{x}))|}}.
\end{gather}

Now, let us consider the columns $\smash{\bar{\psi} \doteq (\bar{\psi}^1, \bar{\psi}^2, \ldots)^\intercal}$ and $\psi \doteq \smash{(\psi^1, \psi^2, \ldots)^\intercal}$ as elements of a vector space $\hV$ with the inner product
\begin{gather}
\braket{\xi|\eta} = g_{ab} \xi^{a*} \eta^b, \quad g_{ab} \doteq \braket{e_a, e_b}.
\end{gather}
(The space $\hV$ is isomorphic to $\hH$ but notice the difference in the notation for the two inner products.) Then, the relation between $\smash{\bar{\psi}^a}$ and $\smash{\psi^b}$ can be expressed in the following vector form:
\begin{gather}\label{eq:Umap}
X: \kpt{5}\psi \mapsto \bar{\psi} = \oper{\mU} \psi,
\quad 
\braket{e^a, \oper{\mU}e_b} = \mU[X]^a{}_b.
\end{gather}
Here, $\oper{\mU}$ is a linear operator whose representation in the basis $\mcu{E}$ is the matrix $\mU^a{}_b$ (and similarly for other operators). The sequence $\psi \mapsto \oper{\mU}\psi \mapsto \oper{\mU}^2\psi \mapsto \ldots \mapsto \oper{\mU}^t\psi$ will be called dynamics or evolution of $\psi$. Accordingly, $\oper{\mU}$ will be called a propagator, and the number of iterations~$t$ will be called time.

The indices in the above formulas can be raised and lowered using $g_{ab}$ as a Hermitian metric \cite{my:wkin}; in particular, \Eq{eq:decomp} can be rewritten as
\begin{gather}\label{eq:decomp2}
\Psi(x) = e^{a}(x)\,\psi_a = e^{a*}(x)\,\psi_a^*
\end{gather} 
(and similarly for $\bar{\Psi}$), where we used that $\Psi$ is real. Then, \Eq{eq:Umn2} can be expressed as 
\begin{gather}\label{eq:UUi}
\mU[X]^a{}_b = \mU[X^{-1}]_b{}^{a*}.
\end{gather}
But note that the transformation induced by $X^{-1}$ is\,\,$\oper{\mU}^{-1}$, so \Eq{eq:UUi} can be rewritten~as
\begin{gather}
\braket{e^a, \oper{\mU}e_b} = \braket{e_b, \oper{\mU}^{-1} e^a}^* = \braket{e^a, (\oper{\mU}^{-1})^\dag e_b}.
\end{gather}
Thus, $\oper{\mU}^\dag = \oper{\mU}^{-1}$, \ie the propagator $\oper{\mU}$ is unitary, and as a corollary,
\begin{gather}\label{eq:cons}
\braket{\bar{\psi}|\bar{\psi}} = \braket{\psi|\oper{\mU}^\dag\oper{\mU}|\psi} = \braket{\psi|\psi}.
\end{gather}

Also note that from \Eqs{eq:decomp} and \eq{eq:decomp2}, the average\footnote{The average $\favr{x}$ over the distribution $F$ in the original classical system \eq{eq:FF} should not be confused with the average \textit{of} $\favr{x}$ over multiple quantum measurements, $\qavr{x}$, performed on a quantum circuit that implements this classical system.} $\favr{x} \doteq \int \dd x\,x\,F(x)$ is naturally understood as a quadratic form evaluated on $\psi$:
\begin{gather}\label{eq:avrx}
\favr{x} = \braket{\psi|\oper{x}|\psi},
\end{gather}
where the (Hermitian) operator $\oper{x}$ is such that its matrix elements in the metric $g_{ab}$ are given by ${x^a}_b = \braket{e^a, x e_b}$. Also assume the notation $\ket{x_\lambda}$ for the eigenvector of $\oper{x}$ corresponding to a given eigenvalue $x_\lambda$; \ie $\oper{x}\ket{x_\lambda} = x_\lambda\ket{x_\lambda}$. If the system is initially in eigenstate $\ket{x_0}$, then $t$ application of the map bring the system to state $\oper{\mU}^t\ket{x_0} \equiv \ket{x_t}$, which is also an eigenstate of $\oper{x}$. Vector sequences $\ket{x_t}$ will be called trajectories generated by~$\oper{\mU}$. (Although $\ket{x_\lambda}$ are fixed, the \textit{label} $\lambda$ evolves with time.)

\section{Truncation and unitarization} 
\label{sec:trunc}

As seen from the previous section, any reversible map, including maps that are nonlinear and non-AP, can be represented as a unitary transformation of a state vector whose components encode the probability distribution. This opens a possibility of computing any reversible map on a QC, provided the map \eq{eq:Umap} can be adequately approximated with a finite-dimensional unitary map (see below). Notably, a quantum implementation will preserve the property \eq{eq:cons} even if the propagator is implemented with a finite accuracy. This is because any implementation of $\oper{\mU}$ on a QC is automatically unitary, as also pointed out in \Refs{ref:georgeot01, ref:georgeot02}. (In contrast, classical calculations are prone to round-off errors and thus are always imprecise, unless they can be done without floating-point operations \cite{foot:zalka}.) However, a practical implementation requires that the infinite-dimensional space $\hV$ be reduced to a subspace $\hV_N$ that has a finite number of dimensions~$N$.

Without loss of generality, let us assume that the basis vectors $e_a$ with $a = 1, 2, \ldots, N$ are a full set of basis vectors for a subspace $\hV_N$. This truncation implies replacing the infinite-dimensional matrix $\mU$ with its upper-left block $\mVN$ of size $N \times N$.  (Here and further, we assume for clarity that the basis $\mcu{E}$ is orthonormal, so the metric $g_{ab}$ is Euclidean; then, upper and lower indices do not have to be distinguished, allowing for a transparent index-free notation.) Unlike $\mU$, the matrix $\mVN$ is generally not unitary as is. To implement its application on a QC, one must approximate $\mVN$ with a different matrix $\mUN$ that \textit{is} unitary and represents a reduced propagator,
\begin{gather}\label{eq:Umapred}
X: \kpt{5}\psi \mapsto \bar{\psi} \approx \oper{\mUN} \psi.
\end{gather}
Such ``unitarization'' can be done automatically as a part of mapping $\mVN$ to a quantum circuit, whatever that procedure may be in a specific case. However, it may be better to do the unitarization controllably, because the choice of $\mUN$ determines the scheme accuracy and complexity. 

Finding a suitable unitarization (which problem is known also, for example, in scattering theory \cite{ref:kowalski65}) can be understood as a closure problem. Its solution is not unique, and the optimal choice of $\mUN$ depends on how one defines the optimum. We present two alternative solutions. First, notice that because $\mU$ is unitary, it is representable as $\mU = \exp(-\ii \mH)$ with Hermitian $\mH$. Suppose one can find $\mH = \ii \ln \mU$; then, one can adopt 
\begin{gather}
H_{ab} = (\mH_{ab} + \mH_{ba}^*)/2, \quad a,b = 1, 2, \ldots, N
\end{gather}
(Hermitization is included in case $\mH$ is known only approximately and thus may not be strictly Hermitian ``out of the box'') and $\mUN = \exp(-\ii H)$, which is unitary because $H$ is Hermitian by construction. This approach is convenient when an analytic formula for $\mH$ is available, like in \Refs{ref:georgeot08, ref:frahm04, ref:levi04, ref:levi03, ref:georgeot01, ref:georgeot01b, ref:benenti01, ref:schack98}. This approach is also convenient when $\mU$ is near-identity (\ie $\mH$ is small), which case is well studied too, specifically, in connection with numerical simulations of Hamiltonian systems \cite{ref:morrison17}. For example, to the leading order, one has
\begin{gather}
H_{ab} \approx (\ii/2)(\mU_{ab} - \mU_{ba}^*), \quad a,b = 1, 2, \ldots, N,
\end{gather}
where $\mU_{ab}$ can be calculated using \Eq{eq:Umn}. 

In many problems, though, finding $\mH$ explicitly is impossible or impractical, so a more general approach is needed. A universal solution can be as follows: (i)~Construct a finite-size matrix $\mVN$ using the corresponding elements of the original infinite-size matrix [\Eq{eq:Umn}],
\begin{gather}\label{eq:mVn}
\mVN_{ab} = \mU_{ab}, \quad a,b = 1, 2, \ldots, N.
\end{gather}
(ii)~Using the singular value decomposition (SVD) \cite{foot:svd}, find the polar decomposition $\mVN = \msf{U} \msf{P}$, where $\msf{U}$ is unitary and $\msf{P} = \smash{(\mVN^{\dag} \mVN)^{1/2}}$ is positive-semidefinite and Hermitian. (iii)~Ignore $\msf{P}$, which is supposed to be close to a unit matrix if the truncated system adequately represents the original system (\ie that for all operators $\oper{A}$ of interest, the averages $\braket{\psi|\oper{A}|\psi}$ in the two systems are approximately the same). (iv)~Adopt $\mUN = \msf{U}$.

Errors introduced by truncation and unitarization can become significant when the map is iterated sufficiently many times. It may not be possible to  robustly estimate these errors without specifying both $X$ and the unitarization procedure; however, let us discuss a particularly notable issue that is generic for implementations of contraction maps. For example, consider a map that has an attracting fixed point at some $x = x_c$. Because the dynamics of the finite-dimensional $\psi$ is by construction reversible, the corresponding distribution cannot approach the attractor indefinitely. Therefore, $\mUN$ necessarily contains elements that will cause scattering of trajectories  away from the fixed point. [Using the extended transformation \eq{eq:exttr} does not help, because the dynamics of $p$ does not affect the dynamics of $x$.] This will lead to spurious echos after some $\tau(N)$ iterations.\footnote{The same pertains to limit cycles and to the extended transformation \eq{eq:exttr}, assuming the system is considered on a finite domain, because the dynamics of $p$ does not affect the dynamics of~$x$.}

The echos can be avoided by using $N$ large enough such that $\tau(N)$ remains larger than the time of interest. The latter would normally be the characteristic time $\mc{T}$ of the system dynamics. It can be estimated by linearizing $X$ in $\xi \doteq x - x_c$ near $\xi = 0$. This leads to $\bar{\xi} = J_c \xi$, where $J_c \doteq J(x_c)$; then, the characteristic time of the dynamics near the fixed point is $\mc{T} \sim |\ln \lambda(x_c)|^{-1}$, where $|\lambda(x_c)| < 1$ is the smallest eigenvalue of $J(x_c)$ (which is assumed contracting here). Similarly, $\mc{T} \sim |\ln \lambda\,|^{-1}$ can be the typical time scale at any $x$, with $\lambda$ being characteristic eigenvalue of $J(x)$ over the whole domain of interest.

The optimal $N$ can be determined numerically by examining the evolution of expectation values that characterize the distribution localization. In some cases, a useful indicator can be the standard deviation $\smash{(\favr{x^2} - \favr{x}^2)^{1/2}}$; however, it is not informative if multiple attracting fixed points are present. A more informative quantity can be the amplitude of a high Fourier harmonic of the classical distribution $F$:
\begin{gather}
\Gamma_k \doteq |\gamma_k|^2, \quad
\gamma_k \doteq \favr{\psi|\ee^{\ii \pi k \cdot \oper{x}}|\psi}.
\end{gather}
(As expectation values of Hermitian operators, the real an imaginary parts of $\gamma_k = \favr{\psi|\cos(\pi k \cdot \oper{x})|\psi} + \ii \favr{\psi|\sin(\pi k \cdot \oper{x})|\psi}$ are measurable separately.) Here $|k|^{-1}$ is roughly the anticipated minimum width of the distribution beyond which spurious echos ensue. On a spatial grid with cell size $\Delta x$, the optimal $\kappa \doteq |k|\Delta x$ should be a small but not-too-small fraction of unity. (For example, a rule of thumb can be to adopt $|\kappa| = 0.1$; see \Sec{sec:example}.) Then, the results of simulations using the truncated matrix can be trusted as long as $\Gamma_k$ remains small (indicating the distribution is far from converging to $x_c$) or grows monotonically (indicating it is converging), and spurious echos due to overconvergence are to be suspected otherwise. We revisit this issue in \Sec{sec:app}.

\section{Sparseness of the unitarized $\boldsymbol{\mUN}$} 
\label{sec:basis}

\subsection{Spatial basis}

The scheme complexity depends on the choice of the basis functions $e_a$. The Carleman and other global bases considered in \Refs{tex:liu20, tex:engel20, tex:lloyd20} may be a good fit \textit{ad~hoc}. However, because the original map is local in $x$, using spatially localized $e_a$ may be more practical when dealing with generic nonlinear problems. For example, one can partition the $x$ space into small elements of size $\Delta x$ and decompose $\Psi$ within each cell using Legendre polynomials up to an order $\ell \ge 0$, with larger $\ell$ leading to higher fidelity of the map representation \cite{ref:wirasaet10}.\footnote{Alternatively, one can expand the configuration space and use the extended transformation \eq{eq:exttr}. Increasing the number of cells in the $p$ subspace at given $x$ serves the same goal as increasing~$\ell$.}

\begin{figure}
\centering
\includegraphics[width=.48\textwidth]{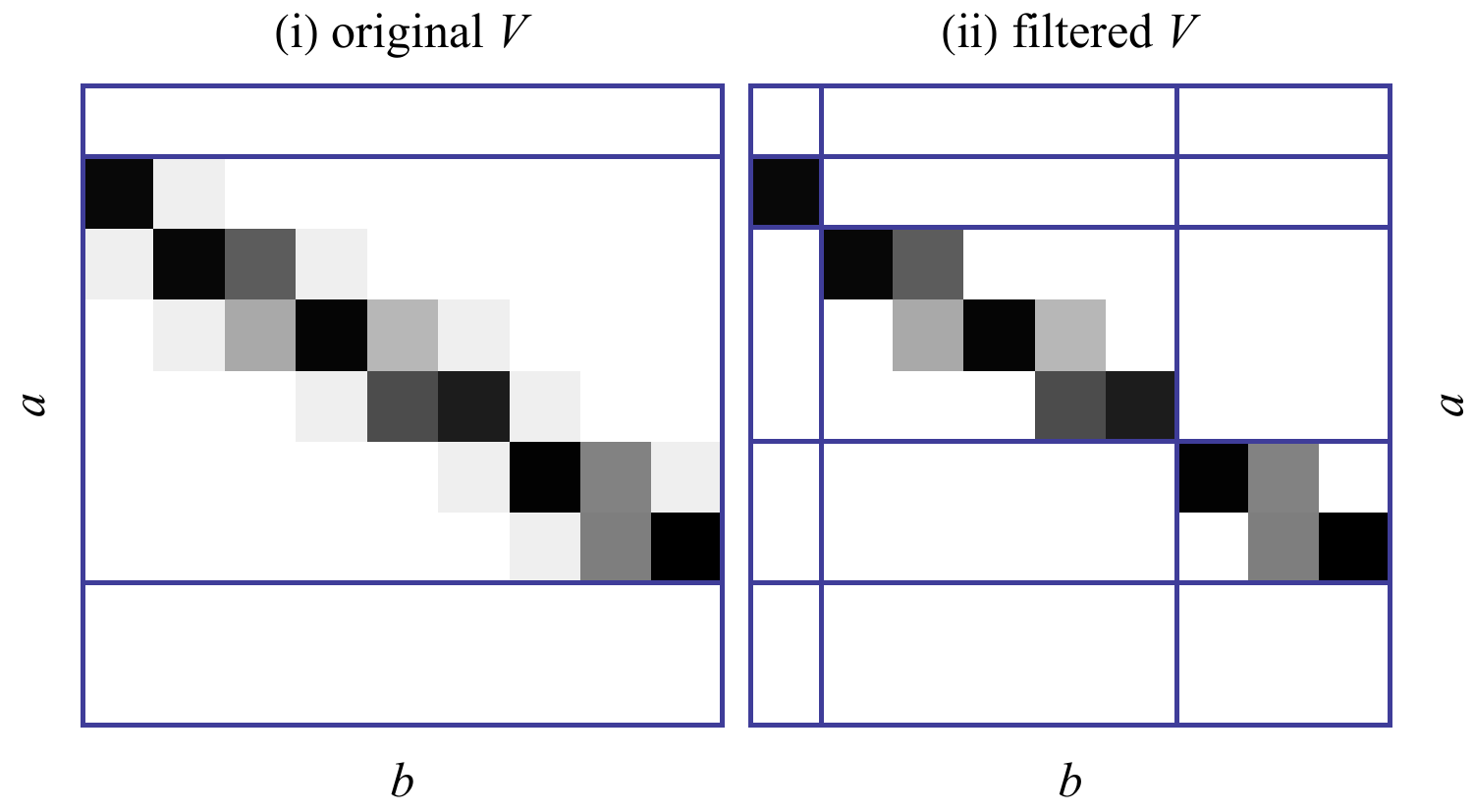}
\caption{A schematic of a typical matrix $\mVN$ in a spatial basis. Shown are $|\mVN_{ab}|$ in the index space $(a, b)$. Non-white elements correspond to nonzero $\mVN_{ab}$. Darker regions correspond to larger $|\mVN_{ab}|$, and blue lines demarcate orthogonal blocks. Shown are: (i)~the original matrix; (ii)~a filtered matrix, where elements with $|\mVN_{ab}| < \epsilon$ are replaced with exact zeros. Unlike the original matrix, the filtered matrix has a block structure, and the typical block size is smaller at larger~$\epsilon$.}
\label{fig:block}
\end{figure}

For simplicity, we limit our consideration to $\ell = 0$. This corresponds to adopting rectangular basis functions that are localized within the corresponding grid cells and constant within these cells. Let us also assume from now on that the mapping is one-dimensional ($n = 1$); then, the assumed basis functions~are
\begin{gather}\label{eq:galb}
e_a(x) = e^{a}(x) = \frac{1}{\sqrt{\Delta x}}\left[1-\Theta
\left(|x-x_a| -\frac{\Delta x}{2}
\right)
\right].
\end{gather}
[Here, $\Theta$ is the Heaviside step function, so each $e_a$ is localized within the corresponding cell $(x_a - \Delta x/2, x_a + \Delta x/2)$.] In such ``spatial'' basis, the (approximate) expressions for $\favr{x}$ and $\gamma_\kappa$, which determines $\Gamma_\kappa \doteq |\gamma_\kappa|^2$,~are
\begin{gather}\label{eq:xg}
\favr{x} = \sum_{a=1}^N x_a |\psi^a|^2,
\quad
\gamma_\kappa = \sum_{a=1}^N \ee^{\ii \pi \kappa a}|\psi^a|^2.
\end{gather}
Also, the nonzero elements $\mVN_{ab}$ of the truncated matrix $\mVN$ in the basis \eq{eq:galb} form a band whose shape in the index space $(a, b)$ is similar to a graph of $X(x)$ up to a vertical flip [\Fig{fig:block}(i)]. At $N \gg 1$, the band is narrow, so $\mVN$ is sparse. However, neighboring columns of $\mVN$ are typically not mutually orthogonal. Unitarization of such a matrix involves orthogonalization of all its columns simultaneously (``global unitarization''), which process generally affects $\mc{O}(N^2)$ matrix elements. This means that unlike $\mVN$, the matrix $\mUN$ generally has $\mc{O}(N^2)$ nonzero elements, \ie may be non-sparse and thus difficult to implement in a quantum circuit. The solution to this is to split $\mVN$ into smaller blocks and unitarize them separately (``local unitarization''), as explained in Secs.~\ref{sec:bs} and \ref{sec:lu}.

\subsection{Block structure}
\label{sec:bs}

Each new block starts in $\mVN$ when the boundary between neighboring grid cells is mapped by $X$ exactly to the boundary between some other cells, \ie $X(x_b + \Delta x/2) = x_a + \Delta x/2$, where $(a, b)$ is a pair of integers. Let us adopt $x_a = (a-1/2)\Delta x$, which can always be done by redefining the origin, and let us measure $x$ in units $\Delta x$, so $\Delta x$ can be omitted. Then, the above equation becomes $X(b) = a$. This means that the fractional part $\zeta(b) \doteq \mathrm{frac}\,b$ of the corresponding $b$ satisfies
\begin{gather}\label{eq:Xab}
\zeta(b) = 0.
\end{gather}

First, let us mention a special case when the map is linear, $X(b) = c + Jb$, $c$ is integer, and $J = r/q$, with $r$ and $q > 0$ being co-primes. In this case, the above equations have a family of exact solutions $(a, b) = (c + rh, qh)$, where $h$ is any integer. Then, $\mVN$ is split into orthogonal blocks of size $r \times q$. If $r \ll N$ and $q \ll N$, this means that the matrix is ready for efficient unitarization (\Sec{sec:lu}).

In general, though, \Eq{eq:Xab} might have no or too few solutions, meaning that $\mVN$ does not split into orthogonal blocks or the blocks are too large. Then, one can reduce the typical block size by zeroing each $\smash{\mVN_{ab}}$ that satisfies
\begin{gather}\label{eq:epsilon}
|\mVN_{ab}| < \epsilon.
\end{gather}
Here, $\epsilon$ is a threshold constant whose optimal value is dictated by how important sparseness is compared to accuracy in a given problem (\Sec{sec:acc}). Then, \Eq{eq:Xab} is replaced with 
\begin{gather}\label{eq:Xab2}
\zeta(b) \lesssim \epsilon.
\end{gather}
For sufficiently large $b$ and generic nonlinear $X$, the left-hand side can be considered as a random number with a flat probability distribution $\mcu{P}(\zeta) = 1$ on the interval $\zeta \in (0, 1)$. This means that the probability of $\zeta$ to be less than $\epsilon$ is $\epsilon$, so it roughly takes $\epsilon^{-1}$ samples of $b$ to satisfy \Eq{eq:Xab2} once. This means that the characteristic block width can be estimated as $\mc{O}(\epsilon^{-1})$. Assuming for simplicity that $J = \mc{O}(1)$, the typical block height (or sparseness of $U$) is also $\mc{O}(\epsilon^{-1})$, and accordingly, the number of blocks is $\mc{O}(\epsilon N)$. 

This readily yields an estimate for the cost of computing $\mUN\psi$. The latter can be represented as a sum over the corresponding blocks, $\smash{\mUN\psi = \sum_i \mUN_i \psi_i}$, where $\psi_i$ are the projections of $\psi$ on the corresponding subspaces. It takes roughly $\epsilon^{-2}$ operations to calculate each $\smash{\mUN_i \psi_i}$ and $\mc{O}(N/\epsilon)$ operations to calculate $\smash{\mUN\psi}$. (By operations, we mean calls of an oracle that encodes elements of $U$.) Computing each next iteration of the map can use the same circuit, in which sense hardware requirements are independent of $t$.

\subsection{Local unitarization}
\label{sec:lu}

Once $\mVN$ has been split into orthogonal blocks, the local unitarization is done as follows.\footnote{This is further discussed in \Sec{sec:example}; see the caption in \Fig{fig:sparse}.} The set of nonzero orthogonal blocks of the filtered matrix $\mVN$ defines a set of subdomains $v_i$ whose images $\bar{v}_i \doteq \mVN v_i$ are mutually orthogonal. Next, consider the set of zero rows of $\mVN$ and the corresponding set of one-dimensional subdomains $\tilde{v}_j$, which is $\ker \mVN$. By combining $v_i$ with  enough $\tilde{v}_j$ and possibly with each other, one can always construct subdomains $w_i$ such that $\dim w_i = \dim (\mVN w_i)$. This is understood as decomposing $\mVN$ into orthogonal \textit{square} blocks $U_i$, albeit parts of these blocks may be disjoint. Such blocks can be unitarized individually; then, the sparseness of $\mUN$ equals the largest $\dim w_i$.

Similar considerations apply to maps on $d$-dimensional grids. Then, $V$ can be considered as a $2d$-dimensional array, and the blocks for local unitarization must be identified in a $2d$ dimensional volume. 

\subsection{Accuracy}
\label{sec:acc}

Since generic $\mVN_{ab}$ are of order $J$, introducing the nonzero threshold value $\epsilon \ll J$ can result in zeroing out only the first and the last band elements per row (\Fig{fig:block}). The follow-up unitarization modifies whole blocks but, by construction, leaves the positive-definite part of $\mUN$ concentrated around the central band. The new nonzero elements that arise at unitarization oscillate rapidly with the index number, so their effect largely averages to zero when the map is applied to a smooth distribution. Hence, spurious dynamics emerges only when the distribution width becomes of the order of the band width (of the original matrix) in the index space, or a few $\Delta x$ in the $x$ space. This means that filtering and unitarization results in shifting trajectories at most by $\mc{O}(\Delta x)$, which will be considered negligible.

\section{Applications} 
\label{sec:app}

Our algorithm for computing a nonlinear map $X$ can be used for: (a) calculating individual trajectories $x(t)$ generated by iterative applications of $X$, and (b) finding attracting fixed points of~$X$, \ie which correspond to $X(x) = x$, or minima of $|X(x) - x|$. Both applications are discussed below. We will use $\qavr{\ldots}$ for the average over quantum measurements of $\oper{x}$; \eg $\qavr{x}$ is the quantum measurement-average approximation to the classical spatial average $\favr{x}$ given by \Eq{eq:xg}, and $\qavr{\Gamma_\kappa}$ is the quantum measurement-average approximation of $\Gamma_\kappa$. The typical number of measurements $M$ sufficient to collect statistics per one measured quantity cannot be estimated \textit{a~priori} without knowing $X$. Thus, measurements must continue until additional measurements stop affecting the averages of interest within the required accuracy.

\subsection{Calculating individual trajectories}
\label{sec:xt}

First, let us discuss how to calculate individual trajectories. In this case, $\psi$ must be sufficiently narrow in the spatial basis. Then $F(t, x) = |e_a(x) \psi^a(t)|^2$ is narrow too, $F(t, x) \approx \delta(x - \favr{x}(t))$. The simulation protocol in this case involves initializing $\psi$, studying the dynamics of $\qavr{\Gamma_\kappa}(t)$ to determine the time interval on which simulations are free from spurious echos (\Sec{sec:trunc}), and finally obtaining $\qavr{x}(t)$ to be used as an approximation of $\favr{x}$.

The specific steps are as follows: (i)~Start with an initial distribution that spans over many grid cells but is narrow compared to the typical scale of $X$.\footnote{If no prior knowledge is available regarding the typical scale of $X$, the whole scheme must be run multiple times, assuming increasingly small scales, until it converges.}  Such a distribution can be prepared efficiently in that $\psi$ is nonzero only in a low-dimensional subspace. (ii)~Pick an integer $t_1 = \mc{O}(\mc{T})$, if $\mc{T}$ can be estimated (\Sec{sec:trunc}), or any $t_1 = \mc{O}(1)$. Apply the map $t_1$ times, then obtain $\qavr{\Gamma_\kappa}(t_1)$. (iii)~Similarly obtain $\qavr{\Gamma_\kappa}$ at $t_2 > t_1$ and $t_3 < t_1$ to determine whether $\qavr{\Gamma_\kappa}$ grows or decreases; find the moment $t_c$ where $\qavr{\Gamma_\kappa}$ has the first local maximum. (iv)~Run the simulation at any desired $t \le t_c$ and obtain $\qavr{x}(t)$; then approximate the classical average $\favr{x}(t)$ as $\favr{x}(t) \approx \qavr{x}(t)$. Provided a reasonable guess on $t_1$, the total number of measurements required for the whole simulation is $\mc{O}(M\mc{T})$.

\subsection{Finding attracting fixed points}

Now let us consider the other potential application, namely, finding attracting fixed points of a given map. If the map of interest is known to have only one attractor, the initial distribution can be arbitrary, and one can proceed as in \Sec{sec:xt}. A more general protocol is as follows: (i)~Start with a flat distribution, which can be prepared efficiently. (ii)~Like in \Sec{sec:xt}, find the moment $t_c$ at which $\qavr{\Gamma_\kappa}$ has a maximum. (iii)~Run the simulation at this $t_c$ and obtain not just $\qavr{x}(t_c)$ but (a discrete approximation to) the \textit{distribution} $F(x, t_c)$ produced by a series of quantum measurements. This distribution will generally have several peaks corresponding to different attractors. The widths of these peaks can be used to infer how close typical trajectories in the corresponding basins of attraction are to the actual attractors. (iv)~If needed, each attractor can then also be explored individually, by placing a narrow packet in its vicinity and proceeding like in \Sec{sec:xt}. 

Modulo multiple measurements, this scheme involves essentially the same calculations as classical optimization algorithms, which trace individual trajectories \cite{book:chong08}. Still, it may have advantage due to simulating \textit{all trajectories at once}. Suppose that $X$ is a black box, so the number and the spatial distribution of attractors is unknown. The quantum scheme on a grid of size $N$ starts with a flat distribution, so it finds all attractors that are separated by at least a few $\Delta x = \mc{O}(N^{-1/d})$, where $d$ is the number of dimensions. A classical scheme that guarantees the same scrutiny necessarily involves tracing $\mc{O}(N)$ trajectories. But this may not be the option if $N$ is extremely large, as can be the case, \eg when $d \gg 1$. In contrast, a quantum scheme can be mapped on $\mc{O}(\ln N)$ qubits, so it is potentially manageable. Of course, calculating the matrix $\mUN$ may be difficult in this case, but this is not a part of a quantum computation \textit{per~se}.

\begin{figure}
\centering
\includegraphics[width=.48\textwidth]{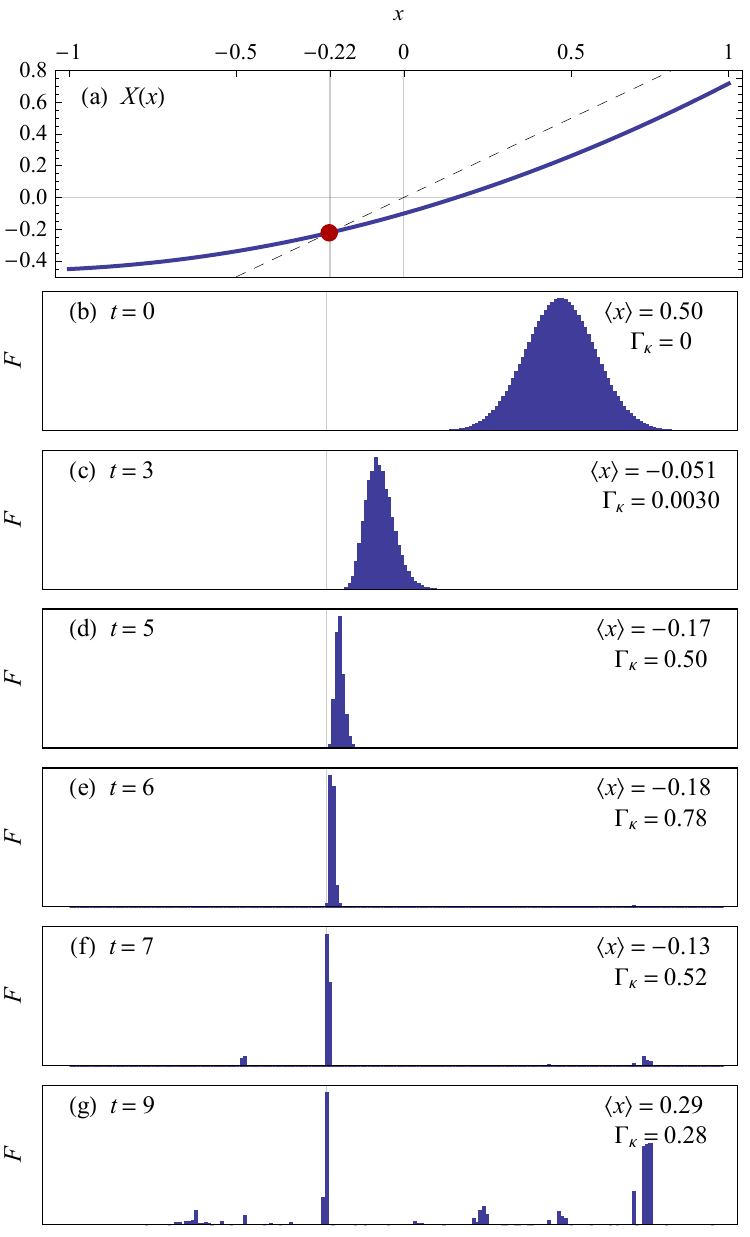}
\caption{(a) The sample map \eq{eq:X} used for simulations (solid blue). The dashed line corresponds to $\bar{x} = x$. The crossing point (red) is an attracting (and the only) fixed point of the map. (b)--(g) The distribution $F(x_a) = (\psi^a)^2$ after $t$ applications of the map $X$ using the matrix from \Fig{fig:sparse}. ($\psi^a$ are exactly real; see the main text.) In each subfigure, $F$ is normalized to its maximum value at the corresponding $t$. [As checked numerically, $\int \dd x\,F(x)$ is the same for all $t$.] In the upper-right corners, shown are the corresponding $\favr{x}$ and $\Gamma_\kappa$ with $\kappa = 0.1$. By $t \sim 6$, the distribution gets localized near the attractor. After that, the errors introduced by unitarization give rise to spurious echos causing delocalization. See also \Fig{fig:x}.}
\label{fig:hist1}
\end{figure}

\section{Example} 
\label{sec:example}

\begin{figure}
\centering
\includegraphics[width=.48\textwidth]{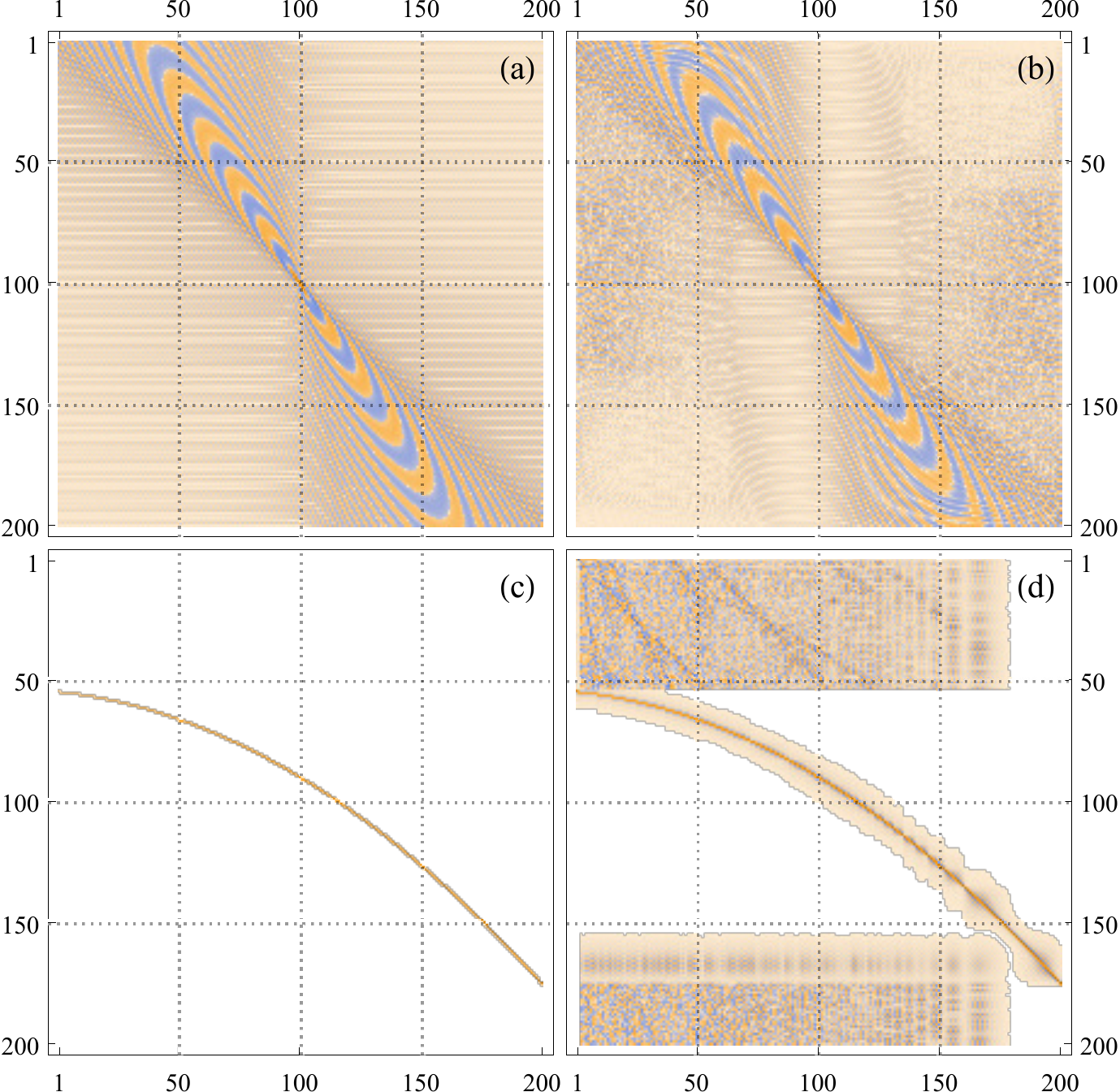}
\caption{Visualization of the matrix structure in the Fourier basis (upper row) and in the spatial basis (lower row): (a, c) the block $\mVN$ of the original infinite-dimensional matrix $\mU$ in the corresponding basis; (b, d) the matrix $\mUN$ obtained by global unitarization of $\mVN$. The axes correspond to the row index and the column index, respectively, like in \Fig{fig:block}. The color intensity reflects the magnitude of the corresponding matrix elements. To improve visualization, only the real part of elements is used, but the imaginary part looks similar.}
\label{fig:matkspace}
\end{figure}

To illustrate our general mathematical scheme, we used finite-width Gaussian $F$ as a proxy for a delta function and, separately, a flat initial $F$. We evolved those on a classical computer with a sample one-dimensional map
\begin{gather}\label{eq:X}
 X(x) = A x^2 + B x + C,
\end{gather}
with $A = 0.25123$, $B = 0.60123$, $C = -0.10123$, and $x \in (-1, 1)$, so that $X$ has a single fixed point $x_c \approx -0.22$, which is an attractor [\Fig{fig:hist1}(a)]. Two orthonormal bases were used, both with $N = 200$. The first one is the Fourier basis that consists of the global complex modes $e_a(x) = e^a(x) = 2^{-1/2}\exp[\ii\pi(a - N/2)x]$. (Real Fourier modes could also have been used, but complex modes allow for a more concise notation.) The second one is the spatial basis \eq{eq:galb} with $\Delta x = 0.01$. In both cases, $\mVN_{ab}$ were calculated using \Eq{eq:Umn}, or explicitly,
\begin{gather}
\mVN_{ab}=\int_{-1}^{1} \dd x\, \sqrt{|X'(x)|}\,e_a^{*}(X(x)) e_b(x).
\end{gather}
The unitarization was done via the polar decomposition \cite{foot:svd}, and at first, a standard subroutine for the SVD was used \cite{foot:math}. [In the spatial basis, $\mVN_{ab}$  are real, so the unitarized matrix is real too by the SVD properties. Then all $\psi^a = \sqrt{F(x_a)}$ remain exactly real when the map is applied.] The corresponding matrices $\mVN$ and $\mUN$ are shown in \Fig{fig:matkspace}. The spatial basis leads to a sparser~$\mVN$, but even in the spatial basis, the corresponding $\mUN$ is not sparse. This is partly due the fact that the coefficients in \Eq{eq:X} are not low-order rationals. 

\begin{figure}
\centering
\includegraphics[width=.48\textwidth]{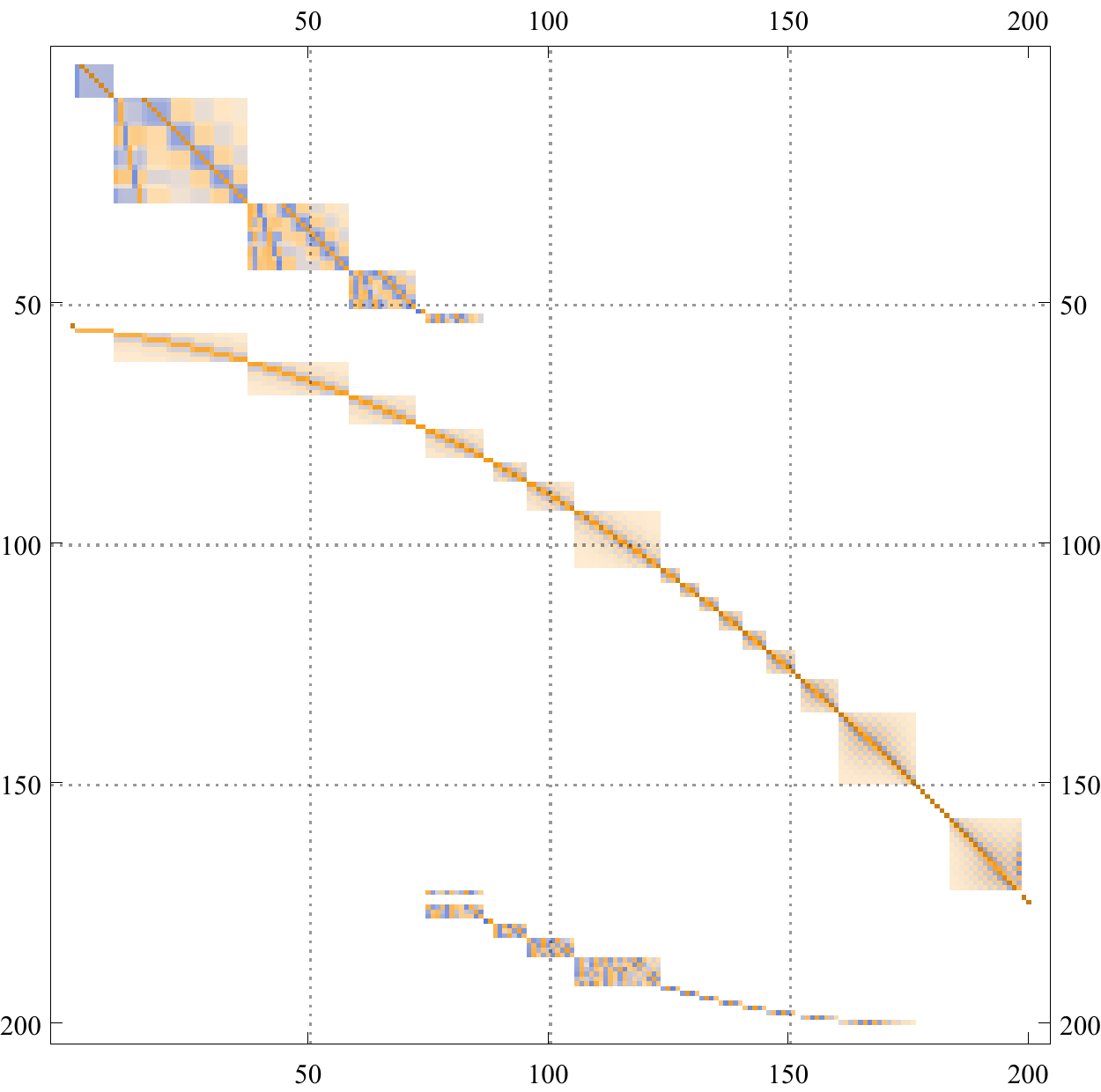}
\caption{Same as in \Fig{fig:matkspace} but for $\mUN$ obtained using block unitarization, namely, as follows: (i)~using $\mVN$ from \Fig{fig:matkspace}(c), construct a filtered $\mVN$ as in \Fig{fig:block}(b) with $\epsilon = 0.1$; (ii)~identify orthogonal blocks $V_i$ of nonzero elements; (iii)~for each $\mVN_i$, construct a block $\smash{\tilde{\mVN}_i}$ of zero rows such that the column $\smash{\{\mVN_i, \tilde{\mVN}_i\}}$ is a square matrix; (iv)~unitarize these square matrices using the polar decomposition; (v)~represent the resulting square blocks columns $\smash{\{\mUN_i, \tilde{\mUN}_i\}}$, where $\mUN_i$ and $\smash{\tilde{\mUN}_i}$ have the same sizes as $\mVN_i$ and $\smash{\tilde{\mVN}_i}$; (vi)~construct $\mUN$ out of $\mVN$ by replacing $\mVN_i$ with $\mUN_i$ and placing $\smash{\tilde{\mUN}_i}$ above or below $\mUN_i$ in rows where the original matrix $\mVN$ is zero (rows with indices $a \lesssim 50$ and $a \gtrsim 170$).}
\label{fig:sparse}
\end{figure}

Using filtering and block unitarization yields a much sparser $\mUN$, as illustrated in \Fig{fig:sparse}. The specific procedure that we used to construct this matrix is described in the figure caption. Notably, $\mUN$ still features nonzero elements in parts where $\mVN$ is zero. However, elements of these blocks oscillate rapidly with the index number, so their effect largely averages to zero when the map is applied to a smooth distribution (as also discussed in \Sec{sec:acc}).

\begin{figure}
\centering
\includegraphics[width=.48\textwidth]{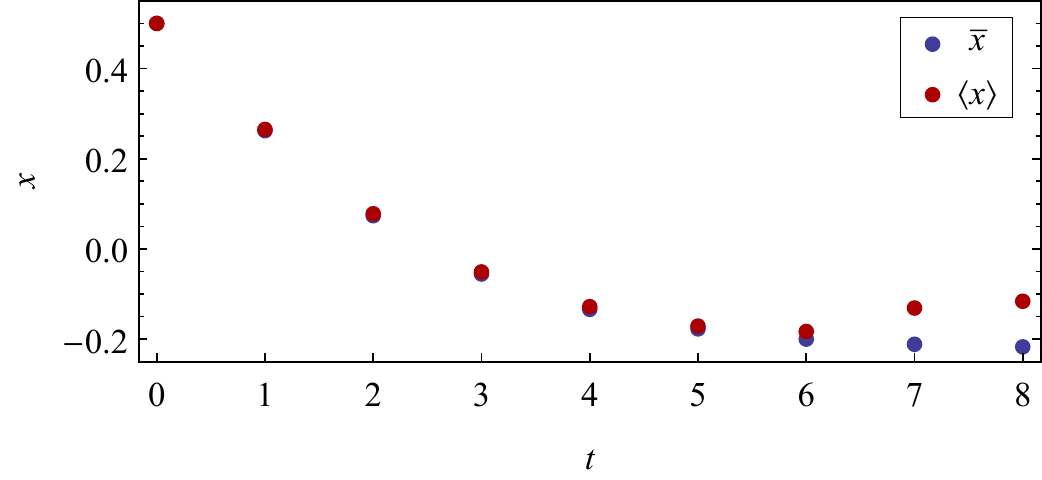}
\caption{Red: $\favr{x}$ from the simulation results shown in \Fig{fig:hist1}. (Simulations with a flat initial distribution yield similar results.) Blue: $\bar{x}$ obtained by $t$ iterations of the map \eq{eq:X} with the initial value $\bar{x}(0) = 0.5$.}
\label{fig:x}
\end{figure}

Simulation results with all three matrices [Figs.~\ref{fig:matkspace}(b), \ref{fig:matkspace}(d), and \ref{fig:sparse}] are similar, so we present the results only for the latter case. Figures \ref{fig:hist1}(b)--(g) show $F$, $\favr{x}$, and $\Gamma_\kappa$ inferred from the vector $\psi$ evolved by $\mUN$ multiple times. For the first few iterations, the reduced unitary map \eq{eq:Umapred} adequately reproduces the true dynamics of the distribution $F$ and $x(t)$, as also seen in \Fig{fig:x}. By $t \sim 6$ [\Fig{fig:hist1}(d)], the distribution gets localized near the attractor, as expected. At later times, the errors introduced by unitarization give rise to spurious echos causing delocalization. The least $|\favr{x} - x_c|$ achieved is $\sim 4\Delta x$, as anticipated, and $\Gamma_\kappa$ with $\kappa = 0.1$ is maximized simultaneously. Other comparable values of $\kappa$ have been tried too and do not change the results significantly. Also, simulations with a flat initial distribution yield similar results (not~shown).

\section{Discussion} 
\label{sec:conc}

In summary, we propose how to compute a general differentiable invertible nonlinear map $x \mapsto X(x)$ on a quantum computer. Our scheme allows calculating individual nonlinear ``trajectories'' $x(t)$ generated by $t$ iterative applications of the map. It can also be applied to minimization algorithms such as Newton’s algorithm and the steepest-descent algorithm. The price of this universality is that the original map is represented adequately only on a finite number of iterations that depends on the resolution. More iterations produce spurious echos, which we propose to monitor via auxiliary measurements. 

In principle, there is another way to avoid echos, which is as follows. Let us consider $y_t \doteq \mVN^t \psi$, where $\psi$ is our usual $N$-dimensional state vector and $\mVN$ is the truncated yet not necessarily unitary matrix \eq{eq:mVn}. The equation for $y \doteq (y_1, y_2, \ldots, y_T)^\intercal$ can be written as
\begin{gather}\label{eq:yV}
\left(
\begin{array}{ccccc}
1 & 0 & \ldots & 0 & 0\\
-\mVN & 1 & \ldots & 0 & 0\\
\vdots & \vdots & \ddots & \vdots & \vdots\\
0 & 0 & \ldots & 1 & 0\\
0 & 0 & \ldots & -\mVN & 1\\
\end{array}
\right)
\left(
\begin{array}{c}
y_1\\
y_2\\
\vdots\\
y_{T-1}\\
y_T
\end{array}
\right)
=
\left(
\begin{array}{c}
\mVN y_0\\
0\\
\vdots\\
0\\
0
\end{array}
\right).
\end{gather}
Because this is a linear equation and the matrix on the left-hand side is sparse, \Eq{eq:yV} can be efficiently solved using well-known quantum algorithms \cite{ref:childs17, ref:harrow09}. This approach does not require unitarization and is analogous to that used in \Refs{tex:liu20, tex:engel20, tex:lloyd20} for solving nonlinear differential equations. However, the problem size is increased from $N$ to $TN$ in this case, and the required circuit becomes much more complicated than an implementation of a single unitary. Also note that this approach can guarantee the smallness only of a relative global error $||\delta y||/||y||$, which is not necessarily of interest. This error can be dominated by contributions from the narrow vicinity of attractors, which says little about the local errors like $|\delta \psi^a|/|\psi^a|$. In contrast, the scheme considered in the previous sections does not have this problem, because it involves no global operations such as matrix inversion.

An in-depth exploration of these issues is left to future work, and so is an explicit implementation of multi-dimensional maps, which we only briefly mentioned above. However, note that the issues considered in this paper will likely be generic to modeling classical dynamics on QCs; for example, they can also be relevant to solving differential equations. Although progress is being made in this area, it is currently restricted to systems with convenient properties, where bounds on errors can be derived explicitly. This luxury will not be an option for simulations of practical interest. For QC to become an applied tool, the focus should be shifted to generic systems, in which case universality and simplicity of algorithms will have priority over guarantees of error bounds. Our work illustrates how challenging it can be to perform even simplest operations on a QC, like computing a nonlinear map, without relying on \textit{ad~hoc} tricks.

The work was supported by the U.S. DOE through Contract No.~DE-AC02-09CH11466.


\begin{thebibliography}{10}
\bibitem{foot:svd}
The SVD of a given matrix $A$ is $A = L D R^\dag$, where $L$ and $R$ are
  unitary and $D$ is a real rectangular diagonal matrix with non-negative
  elements. Adopting $\msf{U} = L R^\dag$ and $\msf{P} = R D R^\dag$ yields the
  polar decomposition~$A = \msf{U} \msf{P}$.
\end{thebibliography}

\end{document}